\documentclass[a4paper,11pt]{article}
\usepackage{jheppub}
\usepackage{color}
\usepackage[dvipsnames]{xcolor}
\usepackage[utf8]{inputenc}

\usepackage{jheppub} 

\def \non {\nonumber}
\def \beq {\begin{equation}}
\def \eeq {\end{equation}}

\DeclareMathOperator{\Tr}{Tr}

\title{\boldmath Kaon Distribution Amplitude from Lattice QCD  and the Flavor SU(3) Symmetry}

\collaboration{Lattice Parton Physics Project (LP${}^3$)}

\author[a]{Jiunn-Wei Chen,}
\author[c, d]{Luchang Jin,}
\author[e, f]{Huey-Wen Lin,}
\author[g]{Andreas Sch\"afer,}
\author[e]{Peng Sun,}
\author[e]{Yi-Bo Yang,}
\author[g]{Jian-Hui Zhang,}
\author[h,i]{Rui Zhang,}
\author[l]{and Yong Zhao}

\affiliation[a]{
Department of Physics, Center for Theoretical Physics, and Leung Center for Cosmology and Particle Astrophysics, National Taiwan University,\\
Taipei, Taiwan 106}

\affiliation[c]{Physics Department, University of Connecticut,\\
Storrs, Connecticut 06269-3046, USA}

\affiliation[d]{RIKEN BNL Research Center, Brookhaven National Laboratory,\\
Upton, NY 11973, USA}

\affiliation[e]{Department of Physics and Astronomy, Michigan State University,\\
East Lansing, MI 48824, USA}

\affiliation[f]{Department of Computational Mathematics, Science and Engineering, Michigan State University,\\
East Lansing, MI 48824, USA}

\affiliation[g]{Institut f\"{u}r Theoretische Physik, Universit\"{a}t Regensburg,\\
D-93040 Regensburg, Germany}

\affiliation[h]{Key Laboratory of Theoretical Physics, Institute of Theoretical Physics,\\
Chinese Academy of Sciences, Beijing 100190, China}
\affiliation[i]{School of Physical Sciences, University of Chinese Academy of Sciences,\\
No.~19A Yuquan Road, Beijing 100049, China}

\affiliation[l]{Center for Theoretical Physics, Massachusetts Institute of Technology,\\
Cambridge, MA 02139, USA}

\emailAdd{jwc@phys.ntu.edu.tw}
\emailAdd{luchang.jin@uconn.edu}
\emailAdd{hwlin@pa.msu.edu}
\emailAdd{andreas.schaefer@ur.de}
\emailAdd{pengsun@pa.msu.edu}
\emailAdd{yangyibo@pa.msu.edu}
\emailAdd{jianhui.zhang@ur.de}
\emailAdd{zhangrui@itp.ac.cn}
\emailAdd{yzhaoqcd@mit.edu}

\abstract{We present the first lattice-QCD calculation of the kaon distribution amplitude using the large-momentum effective theory (LaMET) approach. The momentum-smearing technique has been implemented to improve signals at large meson momenta. We subtract the power divergence due to Wilson line to high precision using multiple lattice spacings.
The kaon structure clearly shows an asymmetry of the distribution amplitude around  $x=1/2$, a clear sign of its skewness. 
We also study the leading SU(3) flavor symmetry breaking relations for the pion, kaon and eta meson distribution amplitudes, and the results are consistent with the prediction
from chiral perturbation theory.}

\begin{document}

\rightline{\sffamily MSUHEP-17-023}

\maketitle
\flushbottom
\allowdisplaybreaks

\section{Introduction}
\label{sec:Introduction}

Meson distribution amplitudes (DAs) $\phi_M$ are important universal quantities appearing in many factorization theorems which {allow for} the description of exclusive processes at large momentum transfers $Q^2 \gg \Lambda^2_\text{QCD}$. Some well-known examples of such processes, which are relevant to measuring fundamental parameters of the Standard Model, include $B \to \pi l \nu, \eta l \nu$ giving the CKM matrix element $ |V_{ub}|$, $B \to D \pi$ used for tagging, and $B \to \pi \pi, K \pi, K\bar{K},\pi \eta, \dots$ which are important {channels} for measuring CP violation (see e.g.~\cite{Stewart:2003gt}). Among those processes, the large difference between the {strength of} direct CP violation for $B^{\pm} \to \pi^0 K^{\pm}$ and $B^{0} \to \pi^{\mp} K^{\pm}$ \cite{Li:2009wba}, and for $D^{0} \to K^+ K^-$ and $D^{0} \to \pi^+ \pi^-$ \cite{Li:2012cfa} clearly highlight the importance of understanding the flavor SU(3) symmetry breaking among light flavors before attributing the effects to enhancement of higher-order amplitudes or even new physics.

In the chiral limit where $m_q \to 0$ with $q=u,d,s$, SU(3) symmetry predicts $\phi_{\pi}=\phi_{K}=\phi_{\eta}=\phi_{0}$. {Away from} the chiral limit, we work in the isospin limit ($m_u=m_d=\bar{m}$) (for simplicity), use the $\overline{\text{MS}}$ scheme, and normalize the {DAs} such that $\int dx\,\phi_{M}(x) = 1$ with meson index $M=\pi,K,\eta$. The leading SU(3) breaking from chiral symmetry takes the form
\begin{align}
\label{eq:phiM}
\phi_M(x,\mu) = \phi_0(x,\mu)+ \sum_{P=\pi, K,\eta} \frac{m_P^2}{(4 \pi f_P)^2}\left[ {\cal E}_M^P(x,\mu) \ln \frac{m_P^2}{\mu_{\chi}^2}+ {\cal F}_M^P(x,\mu,\mu_{\chi}) \right] +\mathcal{O}(m_P^4).
\end{align}
The functions $\phi_0$, ${\cal E}_M^P$ and ${\cal F}_M^P$ are independent of light-quark masses, $f_P$ is the decay constant for meson $P$, $x$ is the fraction of the meson momentum held by the quark, $\mu$ is the factorization scale, and $\mu_{\chi}$ is the dimensional regularization parameter in chiral perturbation theory (ChPT). The $\mu_{\chi}$ dependence in ${\cal F}_M^P$ and  $\ln(m_P^2/\mu_{\chi}^2)$ cancel such that $\phi_M$ is $\mu_{\chi}$ independent. In Ref.~\cite{Chen:2003fp}, it was proven using ChPT that
\begin{align}
{\cal E}_{\pi}^P(x,\mu)={\cal E}_{K}^P(x,\mu)={\cal E}_{\eta}^P(x,\mu)=0
\end{align}
for all $P$. Hence, at $\mathcal{O}(m_q)$, the {DAs} in Eq.~\ref{eq:phiM} are analytic in $m_P^2$, where we have used $m_P^2\propto m_q+\mathcal{O}(m_q^2)$. Ref.~\cite{Chen:2003fp} has also shown that
\begin{align}
\label{eq:phiM1}
\phi_{K^+}(x,\mu)-\phi_{K^-}(x,\mu)= \phi_{K^0}(x,\mu)-\phi_{\bar{K}^0}(x,\mu) \propto m_s - \bar{m}.
\end{align}
Furthermore, at $\mathcal{O}(m_q)$,
\begin{align}
\label{eq:phiM2}
\phi_{\pi}(x,\mu)+3\phi_{\eta}(x,\mu)= 2 \left[ \phi_{K^+}(x,\mu)+\phi_{K^-}(x,\mu) \right]= 2 \left[ \phi_{K^0}(x,\mu)+\phi_{\bar{K}^0}(x,\mu) \right]\,,
\end{align}
and hence,
\begin{align}
\label{eq:phiM2x}
\phi_{\pi}(x,\mu)+3\phi_{\eta}(x,\mu) - 2\phi_{K^+}(x,\mu)-2\phi_{K^-}(x,\mu) = \mathcal{O}(m^2_q).
\end{align}
It will be interesting to investigate whether the above leading SU(3) breaking relations derived from ChPT emerge from direct computations of meson DAs in lattice QCD.

Such direct computations have become possible recently, thanks to {the} large-momentum effective theory (LaMET)~\cite{Ji:2013dva,Ji:2014gla,Ji:2017rah}.
The LaMET method is based on the observation that,
{while in the rest frame of the hadron, parton physics corresponds to lightcone correlations, the same physics can be
obtained through time-independent spatial correlations (now known as quasi-distributions) in the infinite-momentum frame. For finite but large momenta  feasible  in  lattice  simulations,  LaMET can  be  used  to  relate Euclidean quasi-distributions to physical ones through a factorization theorem which involves a
matching and power corrections that are suppressed by the hadron momentum~\cite{Ji:2014gla}.}
In the past few years, there have been many studies on the one-loop matching kernel for the leading-twist PDFs~\cite{Xiong:2013bka,Xiong:2017jtn,Wang:2017qyg,Stewart:2017tvs}, generalized parton distributions (GPDs)~\cite{Ji:2015qla,Xiong:2015nua} and meson DAs~\cite{Ji:2015qla}, as well as on the power corrections~\cite{Lin:2014zya,Alexandrou:2015rja,Chen:2016utp,Radyushkin:2017ffo}.
The renormalization property of quasi-distributions was also investigated~\cite{Ji:2015jwa,Ishikawa:2016znu,Chen:2016fxx,Constantinou:2017sej,Alexandrou:2017huk,Chen:2017mzz,Ji:2017oey,
Ishikawa:2017faj,Green:2017xeu} {with the multiplicative renormalizability established} to all-loop orders. The LaMET approach has been applied to compute the nucleon unpolarized, helicity and transversity PDFs~\cite{Lin:2014zya,Alexandrou:2015rja,Alexandrou:2016jqi,Chen:2016utp,Chen:2017mzz,Lin:2017ani}, as well as the pion DA~\cite{Zhang:2017bzy}.
A first lattice PDF calculation at physical pion mass has recently become available~\cite{Lin:2017ani}.
The $\mathcal{O}(a)$-improved operators associated with large hadron momentum were worked out in Ref.~\cite{Chen:2017mie}.

Motivated by LaMET, it was proposed that one can extract the PDFs from the ``lattice cross sections"~\cite{Ma:2014jla,Ma:2017pxb}, and the quasi-PDF is one of them. More recently, it was suggested that one can study instead an Ioffe-time or pseudo distribution~\cite{Radyushkin:2017cyf} which is related to the quasi-distribution through a simple Fourier transform. While this method shows some interesting renormalization feature~\cite{Orginos:2017kos}, it is essentially equivalent to the LaMET approach~\cite{Ji:2017rah,Izubuchi:2018srq,Zhang:2018ggy} and offers no new physics regarding the factorization into PDFs. In addition, there are proposals using current-current correlators
to compute PDFs, the pion DA, etc.~\cite{Liu:1993cv,Braun:2007wv,Detmold:2005gg,Liang:2017mye,Bali:2017gfr}.
Different approaches can have different systematics to reach the same goal; therefore, they can be complementary to each other.

A first lattice calculation of the leading-twist pion DA using LaMET is done in Ref.~\cite{Zhang:2017bzy}, where the results were improved by a Wilson line renormalization that removes power divergences. 
The plan of the present paper is to extend the study in Ref.~\cite{Zhang:2017bzy} to the $K$ meson and its SU(3) partners, the $\pi$ and $\eta$ mesons. A further improvement was made by implementing the momentum-smearing technique proposed recently~\cite{Bali:2016lva} to increase the overlap with the ground state of a moving hadron. With our results computed directly from lattice QCD, we will examine the ChPT prediction of the leading flavor SU(3) breaking relations in Eqs.~\ref{eq:phiM1} and~\ref{eq:phiM2}.

The rest of the paper is organized as follows: In Sec.~\ref{sec:Methodology} we briefly review the procedure for extracting meson DAs from the quasi-DAs defined in LaMET and {explain how we access the $\eta$ DA}. In Sec.~\ref{sec:lattice_results}, we show {our lattice results. The final result on $\phi_K$ clearly shows its skewness.} Alongside, we have the corresponding results for $\pi$ and $\eta$ mesons and study the leading SU(3)-breaking relations. The conclusion and outlook are given in Sec.~\ref{sec:concl}.

\section{Methodology}
\label{sec:Methodology}

\subsection{Meson DAs from LaMET}
As was explained in Ref.~\cite{Zhang:2017bzy}, in the framework of LaMET, the meson DA
\beq\label{LCDA}
\phi_{M}(x,\mu)={\frac{i}{f_{M}}}\int\frac{d\xi}{2\pi}e^{i(x-1)\xi n\cdot P}\langle M(P)|\bar \psi(0)n\cdot\gamma\gamma_5 \Gamma(0,\xi n)\lambda^a\psi(\xi n)|0\rangle
\eeq
 can be extracted from the quasi-DA
\beq\label{quasiDA}
{\tilde \phi}_M(x,\mu_R, P_z)={\frac{i}{f_{M}}}\int\frac{dz}{2\pi}e^{-i(x-1)P_z z}\langle M(P)|\bar\psi(0)\gamma^z\gamma_5 \Gamma(0,z)\lambda^a\psi(z)|0\rangle,
\eeq
where $\mu$ is a renormalization scale of $\phi_M$ in the $\overline{\text{MS}}$ scheme,
$n^\mu=(1,0,0,-1)/\sqrt 2$ is a lightlike vector, $\Gamma$ is a straight Wilson line that makes the quark bilinear operator gauge invariant,
$\lambda^a=\lambda^3$, $(\lambda^4\pm i \lambda^5)/2$, $\lambda^8$
for $M=\pi$, $K^\pm$, and $\eta$, respectively. In the ${\tilde\phi}_M$ computation,
both the quark bilinear and the meson momentum $P_z$ are along the $z$ direction.
$\mu_R$ denotes the renormalization scale of ${\tilde \phi}_M$ in a given scheme.
After removing the power corrections, ${\phi}_M$ and ${\tilde\phi}_M$ are the same in the infrared. Their difference in the ultraviolet can be compensated by the matching kernel $Z_\phi$, which can be computed perturbatively~\cite{Ji:2015qla}:
\beq\label{pionDA1loopmatching}
{\tilde \phi}_M(x, \mu_R, P_z)=\int_0^1 dy\, Z_\phi(x, y, \mu, \mu_R, P_z)\phi_M(y, \mu)+\mathcal{O}\left(\frac{\Lambda^2_\text{QCD}}{P_z^2},\frac{m^2_M}{P_z^2}\right).
\eeq
The matching kernel $Z_\phi$ has the form
\begin{align}\label{matching}
Z_\phi(x, y) &= \delta (x-y) + \frac{\alpha_s}{2\pi} \overline{Z}_\phi(x, y)
  + \mathcal{O}\left(\alpha_s^2 \right)\non\\
&=\delta(x-y)+\frac{\alpha_s}{2\pi}\big(Z_\phi^{(1)}(x, y) - C\delta(x-y)\big)+ \mathcal{O}\left(\alpha_s^2 \right)
\end{align}
with $C=\int_{-\infty}^{\infty} d x'\,Z_\phi^{(1)}(x', y)$. The expression for $Z_\phi^{(1)}(x, y)$ can be found in Ref.~\cite{Zhang:2017bzy}.
Eq.~\ref{matching} tells us that $\phi_M$ and ${\tilde\phi}_M$ differ only at loop level, thus we can write (ignoring the power corrections for the moment)
\begin{align}
\phi_M(x) &\simeq \tilde{\phi}_M(x)
  - \frac{\alpha_s}{2\pi} \int dy\,
    \overline{Z}_\phi\!\left(x, y\right)\tilde{\phi}_M(y)\non\\
&\simeq \tilde{\phi}_M(x)
  - \frac{\alpha_s}{2\pi} \int_{-\infty}^{\infty}\!dy\,
  \left[ Z_\phi^{(1)}\!\left(x, y\right)
    \tilde{\phi}_M(y)
  - Z_\phi^{(1)}\!\left(y, x\right)
    \tilde{\phi}_M(x)\right]
\end{align}
with an error of $\mathcal{O}\left(\alpha_s^2\right)$~\cite{Ma:2014jla}. For simplicity, we have also extended the integration range of $y$ to infinity, which will introduce an error at higher order.
To account for the power corrections, we need to know higher-twist and meson-mass corrections as well. The meson-mass corrections have been computed to all orders in $m_M^2/P_z^2$~\cite{Zhang:2017bzy}, while the higher-twist corrections were removed by a simple fitting with a polynomial form in $1/P_z^2$. In this work, we will follow the same procedure but leave out the higher-twist corrections,
because we have observed non-monotonic behavior in $P_z$ in our lattice data. This implies that the polynomial fit might be too naive to account for the higher-twist effects.

\subsection{Accessing the Meson DA Matrix Element on the Lattice}

We start from the calculation of the correlator,
\begin{align}
\label{eq:2pt}
\tilde{C}(z, P_z, \tau) = \left\langle \int d^3 x\, e^{i \vec{P}\cdot x} \bar{\psi}(\vec{x},\tau) \gamma^z\gamma_5\Gamma(\vec{x},{\vec{x}+z} ){\lambda^a}^{\dagger}\psi(\vec{x}+z,\tau)\  \bar{\psi}^S(0,0) \gamma_5 \lambda^a \psi^S(0,0)\right\rangle,
\end{align}
where the sink operator at timeslice $\tau$ is the Fourier transform of the quasi-DA, and the quark fields $\psi^S$ in the source operator at timeslice $0$ {have been momentum smeared}~\cite{Bali:2016lva},
\begin{align}
\psi^S(x)= \int d^3 y\, e^{-\frac{|x-y|^2}{2\sigma^2}-i\vec{k}\cdot{(\vec{x}-\vec{y})}}U(x,y)\psi(y),
\end{align}
where $U(x,y)$ is the gauge link that makes $\psi^S(x)$ gauge covariant, $\sigma$ is the smearing {radius}.
Following Ref.~\cite{Bali:2016lva}, the momentum smearing parameter $\vec{k}$ is determined by optimizing the signal of $\tilde{C}(z, P_z, \tau)$. We found that $\vec{k}=\pm 0.73\vec{P_z}$ for the quark(antiquark) is suitable for our calculations with $\vec{P_z}=(0,0,\{4,6,8\}\pi/L)$. Note that we need to generate the quark and antiquark propagators separately, since the optimal $\vec{k}$'s for them have opposite {signs}.

Following the standard procedure, we insert a complete set of states between the two operators at timeslices $t$ and $0$ in $\tilde{C}$ Eq.~\ref{eq:2pt}. Then, assuming the complete set of states is saturated by the ground state of energy $E_0$
and an effective excited state of energy $E_1$
 at large $t$, we have
\begin{align}
\label{C2x}
\tilde{C}(z, P_z, \tau)= \frac{Z_\text{src} \tilde{h}_M(z,P_z)}{2 E_0} e^{-E_0\tau}+B(z,P_z) e^{-E_1\tau}
\end{align}
with the matrix element
\begin{align}
\label{hb0}
\tilde{h}_M(z,P_z)=\langle M(P)|\bar\psi(0)\gamma^z\gamma_5 \lambda^a\Gamma(0,z)\psi(z)|0\rangle .
\end{align}
What we need for the quasi-DA calculation is the normalized $\tilde{h}$ defined as
\begin{align}
\label{hb}
h_M(z P_z, P_z)=\frac{\tilde{h}_M(z,P_z)}{P_z f_M} ,
\end{align}
which satisfies $h_M(0,P_z)=1$. Therefore, even if we do not separate $\tilde{h}_M$ from the $z$-independent source matrix element $Z_\text{src}$, {the determination of $h_M$ is not affected}.

\subsection{Accessing the $\eta$ Distribution Amplitude}
\label{subsec:eta}

For $\pi$ and $K$, $\tilde{C}$ in Eq.~\ref{eq:2pt} receives contributions from connected diagrams only.  For $\eta$, in addition to connected diagram contributions, $\tilde{C}$ also receives contributions from disconnected diagrams. However, the disconnected diagram is $\mathcal{O}((m_s-\bar{m})^2)$ suppressed because there are two fermion loops; each of which is suppressed by one power of $(m_s-\bar{m})$ in the diagram. Therefore, it seems that if we just work at $\mathcal{O}(m_q)$, we can safely neglect the disconnected diagram of $\eta$.

However, by dropping the disconnected diagrams, the $u(d)$ and $s$ quark contributions in $\tilde{C}$ yield different values of ground-state energy $E_0$; that is, $\Delta E_0 \equiv E_0^s-E_0^{u(d)} \neq 0$. Then, when $\tau > 1/|\Delta E_0|$, $\tilde{C}$ is dominated by the quark contribution {of} lower $E_0$. However, when the hadron momentum $P_z$ is large, such that $E_0 \simeq P_z + m_{\bar{q}q}^2/2 P_z$, where $m_{\bar{q}q}$ is the mass of the $\bar{q}q$ state without the disconnected diagram, {as} long as the plateaus for the mass determination appear within
$\tau < 2 P_z/| m_{\bar{s}s}^2- m_{\bar{u}u}^2|$, $\tilde{C}$ remains equally balanced between $u(d)$ and $s$ quark contributions. Therefore, even without including the disconnected diagrams, the error from this ambiguity can be systematically reduced by increasing $P_z$.

There is another complication for $\eta$. That is, the operator associated with $\lambda^8$ creates the $\eta_8$ meson. But the physical $\eta$ is a linear combination of  $\eta_8$ and $\eta_0$, the SU(3) singlet. Fortunately, the mixing angle $\theta$
is small ($\theta \approx -15^\circ$) \cite{Feldmann:1998vh}. Therefore, in Eq.~\ref{eq:2pt}, when we insert the physical $\eta$ state between the two operators at time slices $\tau$ and $0$, the $\eta_0$ contribution is suppressed by a mixing factor
$\sin \theta \approx 0.08$ times a factor of $(m_s-\bar{m})$ coming from the overlap of $\eta_0$ {with} the $\lambda^8$ type operator. Hence, numerically, the  $\eta_0$ contribution can be counted as $\mathcal{O}((m_s-\bar{m})^2)$ and can be neglected in our calculation.

The above discussion leads to the conclusion that if the plateaus for the meson-mass determination appear within
$\tau < 2 P_z/| m_{\bar{s}s}^2- m_{\bar{u}u}^2|$, then the connected diagram contribution of $\tilde{C}$ of Eq.~\ref{eq:2pt} with $\lambda^a=\lambda^8$ yields contributions from $u,d,s$ quarks in the ratio $1:1:4$. This $\tilde{C}$ can be determined from Eq.~\ref{C2x} with the matrix element of Eq.~\ref{hb} associated with the $\eta_8$ DA. This implies that for the largest $P_z$ we use ($8\pi/L$), the plateaus should be reached within $\tau<1/|\Delta E| \simeq 20 a$, which is clearly satisfied.

In the following sections, we first present the unphysical $\eta_s$ results (with connected-diagram contributions only) for different $P_z$ values. Then, based on the above discussion,
we approximate the $\eta_8$ DA for $P_z=8\pi/L$ with $(\phi_{\pi}+2\phi_{\eta_s})/3$ computed using connected diagrams only, and use the result to check the SU(3) relation Eq.~\ref{eq:phiM2}.

\section{Lattice results}
\label{sec:lattice_results}

In this section, we present our lattice setup and the results for the $\pi$, $K$ and $\eta$ DAs.
The simulations were performed using clover valence fermions on a $24^3\times 64$ lattice with 2+1+1 flavors (degenerate up and down, strange, and charm degrees of freedom) of highly improved staggered quarks (HISQ)~\cite{Follana:2006rc} generated by the MILC Collaboration~\cite{Bazavov:2012xda}. The pion mass on this ensemble is 310~MeV, and the lattice spacing $a\approx 0.12$~fm.
In this work, hypercubic (HYP) {smearing~\cite{Hasenfratz:2001hp} is applied to the configurations}
; the bare quark masses and clover parameters are tuned to recover the lowest pion mass of the staggered quarks in the sea.
The results shown in this section were obtained using the correlators calculated from 3 momentum-smearing sources and 4 source locations on each of the 967 configurations.

\begin{figure}[htbp]
\centering 
\includegraphics[width=.48\textwidth,]{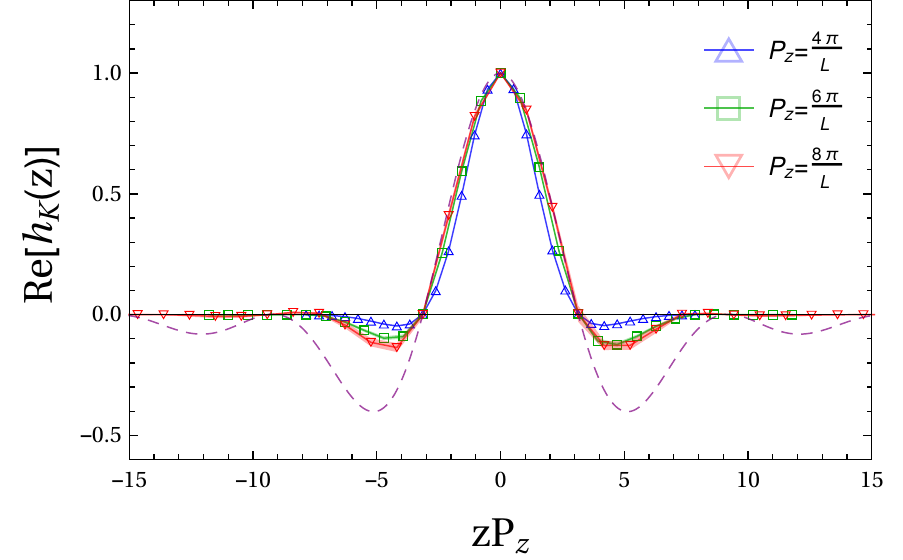}
\includegraphics[width=.48\textwidth,]{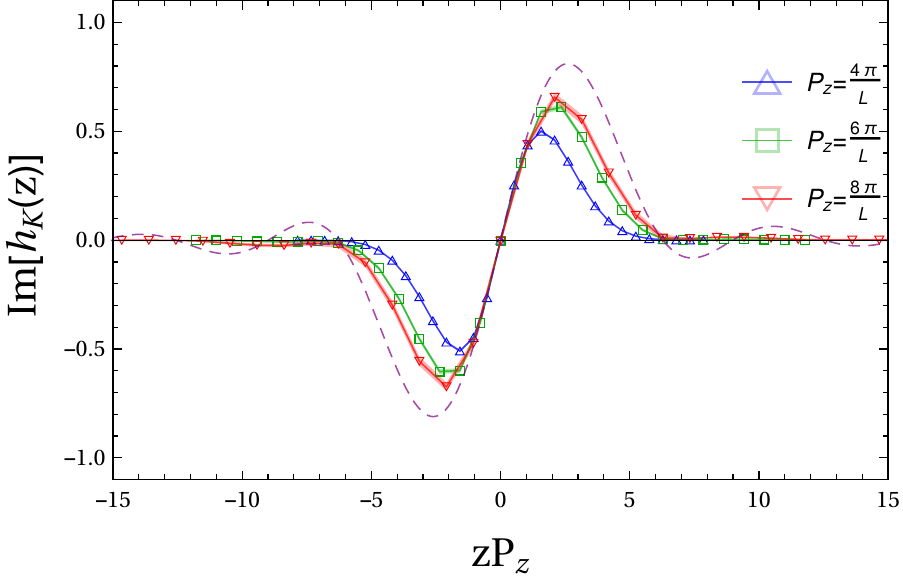}
\includegraphics[width=.48\textwidth,]{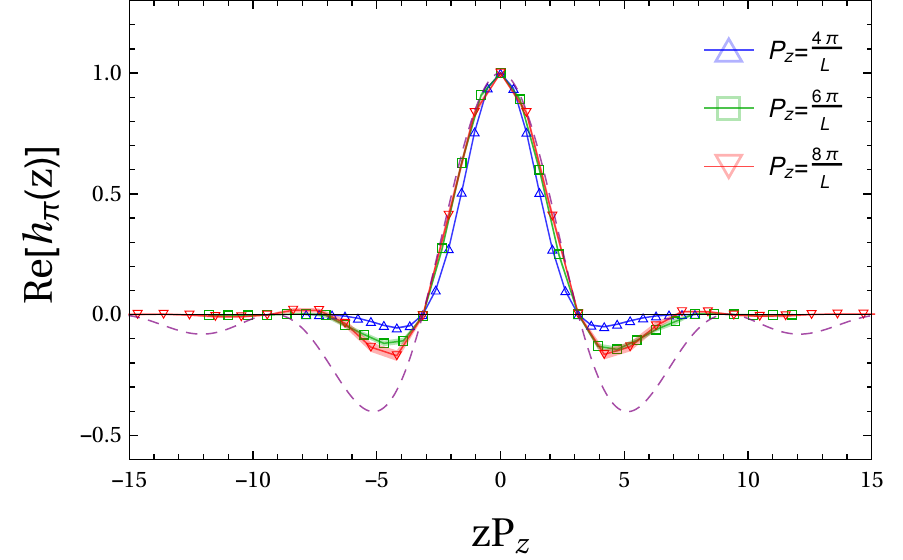}
\includegraphics[width=.48\textwidth,]{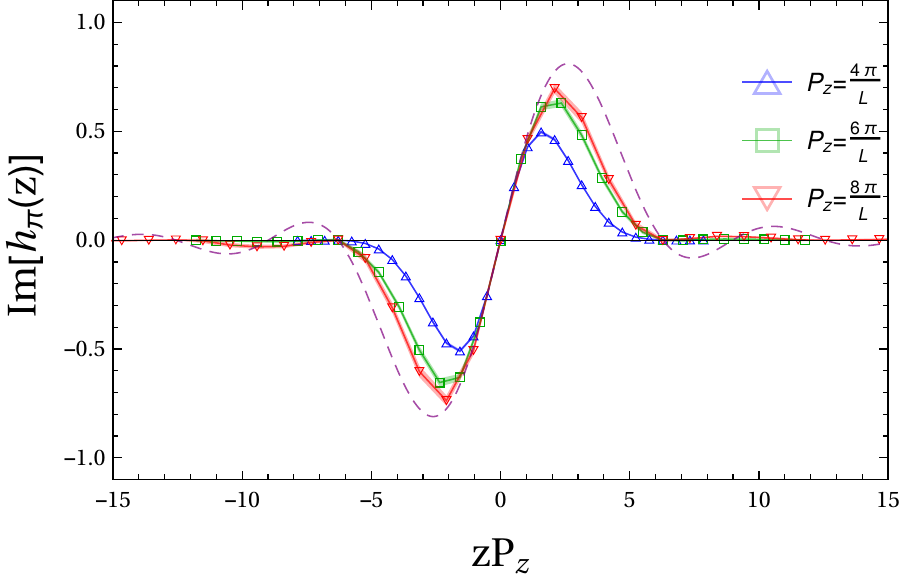}
\includegraphics[width=.48\textwidth,]{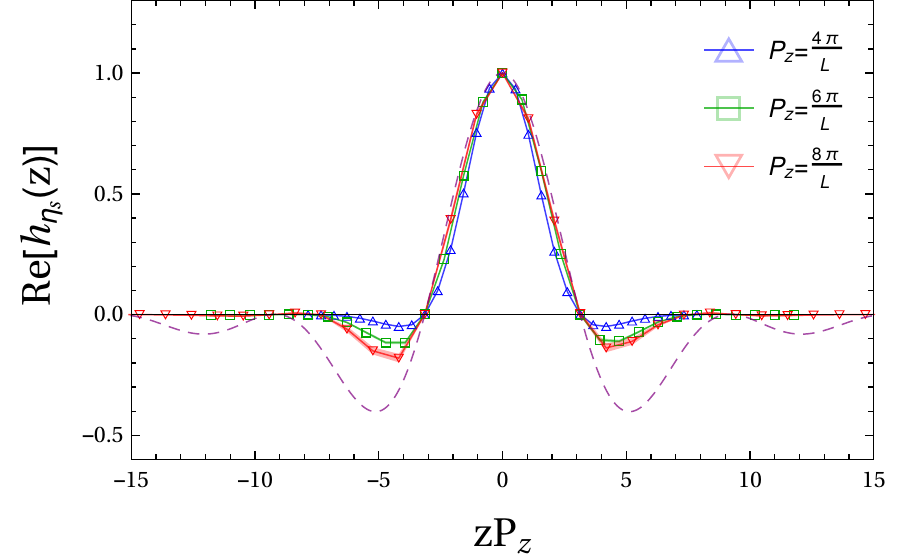}
\includegraphics[width=.48\textwidth,]{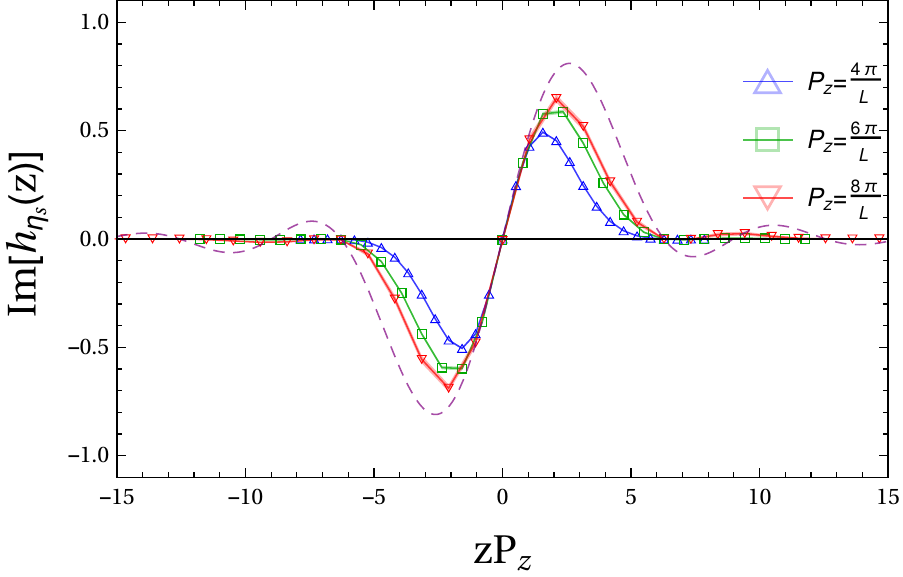}
\caption{\label{fig:quasiDAmatrix}
The quasi-DA matrix elements $h_M(z, P_z)$ for $M=K$ (top), $\pi$ (middle) and $\eta_s$ (bottom), respectively,
shown as functions of $z P_z$ at three different values of $P_z$. Note that $h_{\eta_s}$ contains the connected diagram contributions only.
The purple dashed lines are Fourier transforms of the asymptotic DA $\phi(x)=6x(1-x)$.
}
\end{figure}

The bare matrix elements $h_M$ defined in Eq.~\ref{hb} for kaon, pion and $\eta_s$ ({with connected diagrams only}) are shown in Fig.~\ref{fig:quasiDAmatrix} for $P_z=4\pi/L, 6\pi/L$, and $8\pi/L$.
The dispersion relation, $E_0^2(P_z)=m^2+\hat{P}_z^2$ with $\hat{P}_z=2\sin({P_z}/{2})$, is satisfied up to two standard deviations within statistical uncertainties {for all the $P_z$'s used in this work}, ruling out sizable systematics due to discretization. The Fourier transform of the DA asymptotic form, $\phi(x)=6x(1-x)$,
 is also shown in Fig.~\ref{fig:quasiDAmatrix}. If {the} asymptotic form is correct, it suggests one needs to push for even {larger} $zP_z$ to catch all the short distance features; for the present calculation, this translates into a meson DA with less precision in the regions near $x=0$ and $x=1$.
However, the asymptotic DA is defined in another renormalization scheme, so this is not a direct comparison with $h_M$.
We do observe similar oscillating behaviour in our data. {When we increase $P_z$}, the secondary peaks become more pronounced. But the difference between $P_z=6\pi / L$ and $8\pi / L$ is small already.
Nevertheless, we plan to repeat this work with larger boost momentum (to extend the $zP_z$ reach) and reduce the lattice spacing by at least a factor of 2 in the future.

\subsection{Improved Distribution Amplitude}
\label{subsec:improved}

With the DA matrix elements $h_M$, we can then Fourier transform according to Eq.~\ref{quasiDA} to study the meson DAs. To cancel the power divergence in $h_M$ arising from the Wilson-line self-energy diagrams
 we introduce a counterterm $\delta m$, as suggested in Refs.~\cite{Chen:2016fxx,Zhang:2017bzy},
 such that the matching kernel $Z_\phi$ only has logarithmic divergence but no power-divergent contributions.
Thus, the ``improved'' meson quasi-DA~\cite{Zhang:2017bzy} is
\begin{align}
\label{eq:improvedQuasiDA}
\tilde{\phi}^\text{imp}_{M}(x,P_z) &= \int^{\infty}_{-\infty} \frac{d z}{2 \pi}e^{-i(x-1)z P_z+\delta m |z|} P_z h_M(z,P_z).
\end{align}
We then apply the matching kernel $Z_\phi$ and mass correction, as discussed in Sec.~\ref{sec:Methodology}, to obtain the final DA.

First, we need to calculate the counterterm $\delta m$.
The Wilson line can be equivalently described by a quark propagator in the heavy-quark limit
and the only dimensionful counterterm in the heavy-quark Lagrangian is the mass counterterm $\delta m$. Therefore, $\delta m$ can be determined by the Wilson loop $W(\tau,r)$ with width $r$ and length $\tau$, which has the negative effective action of a static quark-antiquark pair with interquark distance $r$ at temperature $1/\tau$.
The quark-antiquark effective potential is approximated by
\begin{equation}
\label{V}
V(r)= -\frac{1}{a}\lim_{\tau\to \infty} \ln\frac{\langle\Tr[W(\tau,r)]\rangle}{\langle \Tr[W(\tau-a,r)]\rangle},
\end{equation}
using a combination of such Wilson loops.
The cusp anomalous dimensions from the four sharp corners of the Wilson loop are canceled between the numerator and denominator of the expression, and keeping $1/\tau$ larger than the inverse of the energy gap between the ground state and the first excited state ensures that higher excitations are sufficiently suppressed.

When $r$ is larger than the confinement scale but shorter than the string-breaking scale, this can be fit by
\begin{equation}
\label{Vc}
V(r)=\frac{c_{-1}}{r}+c_0+c_1 r,
\end{equation}
where the $c_{-1}$ term is the Coulomb potential that dominates at short distance and the $c_1$ term is the confinement linear potential. $c_0$ is of mass dimension one, so we can break it into a divergent piece and a finite one in the continuum limit: $c_0=c_{0,1}/a+c_{0,2}$. Then
\begin{equation}
  \delta m={-}\frac{c_{0,1}}{2a},
\end{equation}
where the $2$ compensates for the potential using a quark-antiquark pair.

\begin{figure}[tbp]
\centering 
\includegraphics[width=.55\textwidth,]{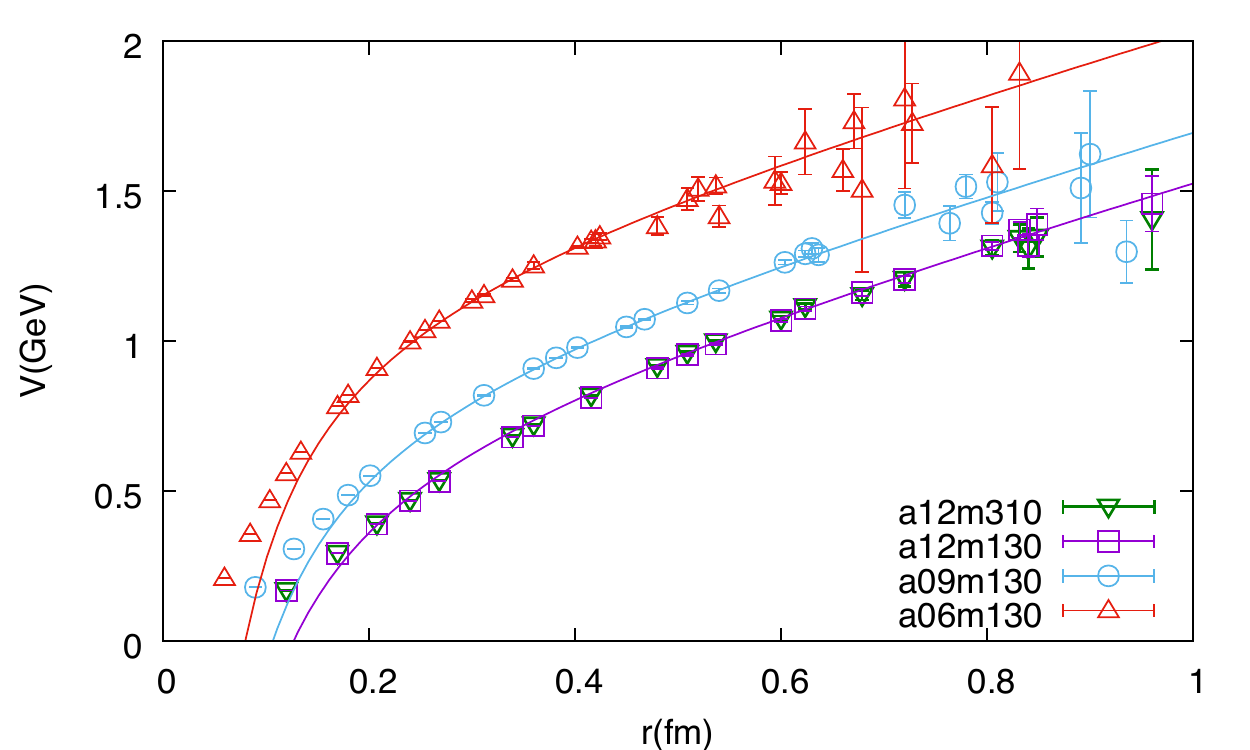}
\caption{\label{fig:potential}
The potential between a static quark and a static antiquark as function of distance $r$ (in fm) computed using Eq.~\ref{V} for $a=0.06$~fm (triangles), 0.09~fm (circles), and 0.12~fm (squares) with pion mass around 135~MeV and for heavier pion mass 310~MeV at $a=0.12$~fm (inverted triangles). The difference between pion masses is negligible for 310 and 135~MeV at 0.12-fm lattice spacing.
The line indicates our attempted fit to the potentials  using {the} form of Eq.~\ref{Vc} with $r\ge 5a$ to extract {the} $\delta m$ counterterm, and the fit describes the majority of the potential well.
}
\end{figure}

Fig.~\ref{fig:potential} shows the effective potential $V(r)$ at lattice spacings $a=0.06, 0.09, 0.12$~fm for $M_\pi=130$~MeV and $a=0.12$~fm for $M_\pi=310$~MeV. The $M_\pi$ dependence for $a=0.12$-fm ensembles is almost undetectable. {A fit of the potential with four parameters $c_{-1}, c_{0,1}, c_{0,2}$ and $c_{1}$, $V(r\ge 5a)$ has a very good} $\chi^2/\text{d.o.f.}=1.04$ (46 degrees of freedom). This fit yields $\delta m=0.154(2)/a$,
which corresponds to $253(3)$~MeV at $a=0.12$~fm.

With the thus determined $\delta m$, we can now obtain $\tilde{\phi}^\text{imp}_{M}(x,P_z)$ for each meson using Eq.~\ref{eq:improvedQuasiDA}.
Next, we apply the one-loop matching kernel $Z_\phi^{(1)}$ (see Eq.~A.6 of Ref.~\cite{Zhang:2017bzy}), which is essential in LaMET to obtain lightcone quantities from the quasi-distribution
\begin{align}
\phi^{\text{imp},\text{match}}_M(x, P_z)
&\simeq \tilde{\phi}^\text{imp}_M(x,P_z)
  - \frac{\alpha_s}{2\pi} \int_{-\infty}^{\infty}\!dy\,
  \left[ Z_\phi^{(1)}\!\left(x, y\right)
    \tilde{\phi}^\text{imp}_M(y,P_z)
  - Z_\phi^{(1)}\!\left(y, x\right)
    \tilde{\phi}^\text{imp}_M(x,P_z)\right]
\end{align}
with an error of $\mathcal{O}\left(\alpha_s^2\right)$~\cite{Ma:2014jla} as discussed earlier in Sec.~2.1.
We then apply the mass corrections~\cite{Zhang:2017bzy} to $\phi^{\text{imp},\text{match}}_{M}$ to get the final DAs
\begin{align}
\label{eq:masscorr}
\phi_{M}=\sqrt{1+4c}\sum_{n=0}^\infty \frac{(4c)^n}{f_+^{2n+1}}\Bigg[&(1+(-1)^n)\phi^{\text{imp},\text{match}}_{M}\Big(\frac{1}{2}-\frac{f_+^{2n+1}(1-2x)}{4(4c)^n}\Big)\non\\
{}+{}&(1-(-1)^n)\phi^{\text{imp},\text{match}}_{M}\Big(\frac{1}{2}+\frac{f_+^{2n+1}(1-2x)}{4(4c)^n}\Big)\Bigg],
\end{align}
where $c=m_M^2/4P_z^2$ and $ f_{+}=\sqrt{1+4c}+1$.
The remaining higher-twist effect is {of $\mathcal O({\Lambda^2_\text{QCD}}/{P_z^2})$}, which is small at our largest 2 momenta used in this work.

\subsection{Kaon Distribution Amplitude}
\label{subsec:Kaon}

\begin{figure}[tbp]
\centering 
\includegraphics[width=.48\textwidth,]{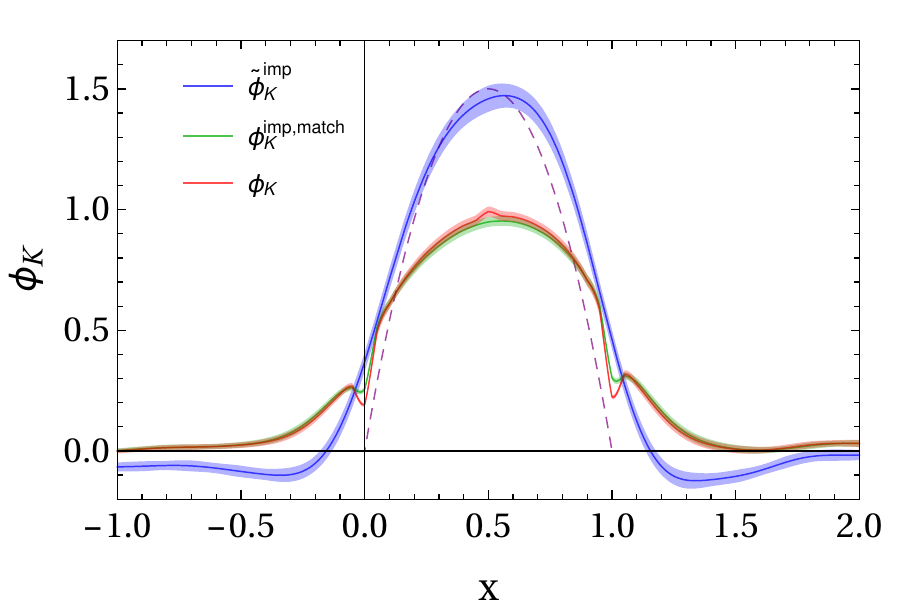}
\includegraphics[width=.48\textwidth,]{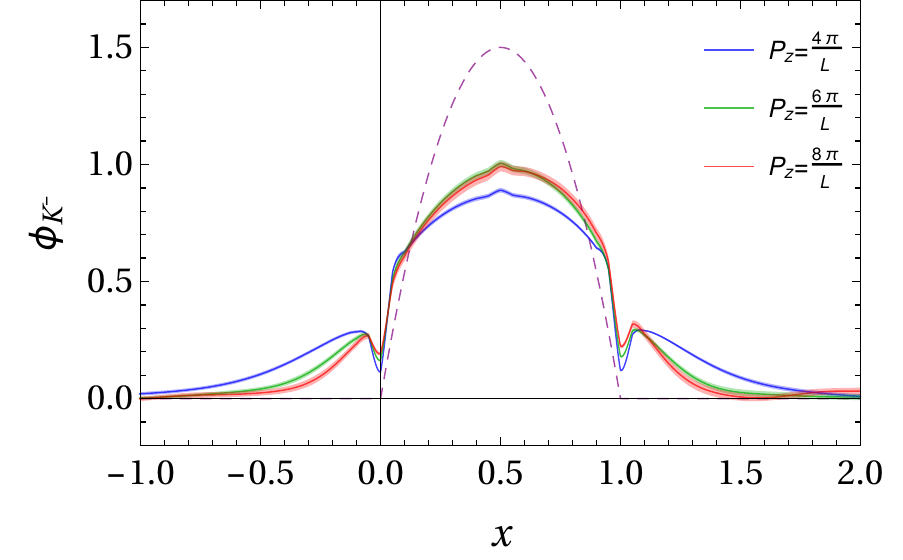}
\caption{\label{fig:Kaon}
(Left) Improved kaon quasi-DA, $\tilde{\phi}^\text{imp}_K(x)$ (blue), kaon DA with one-loop matching applied, ${\phi}^{\text{imp},\text{match}}_K$ (green), {and $\phi_K$ (red) with the meson-mass correction, which is a very small effect, further added.} The asymptotic DA is shown as the purple dashed line.
(Right) DAs for $K^-$ after the one-loop matching and mass corrections but not higher twist corrections are shown with each $P_z$ we study in this work.
}
\end{figure}

Let us now consider the first results for the kaon DA from lattice QCD. The left-hand side of
 Fig.~\ref{fig:Kaon} shows a comparison of $\tilde{\phi}^\text{imp}_{K}(x)$ (improved quasi-DA shown in blue), $\tilde{\phi}^{\text{imp},\text{match}}_{K}(x)$ (after matching quasi-DA to lightcone DA shown in green) and $\phi_K(x)$ (DA with meson mass correction {added}, shown in red) from our largest meson momentum.
The distribution after applying the one-loop matching $\tilde{\phi}^{\text{imp},\text{match}}_{K}(x)$ changes quite significantly from the quasi-distributions.
Further treatment with the meson mass correction yields $\phi_K(x)$ which is very close to $\tilde{\phi}^{\text{imp},\text{match}}_{K}(x)$. This is expected with the large momentum used here.

The right-hand side of Fig.~\ref{fig:Kaon} shows the momentum dependence of $\phi_K(x)$.
Note that the higher-twist correction is not extrapolated away as in our previous work due to the non-monotonic behavior in $P_z$. However, we expect its effect to be small at the largest two momenta used in this work, since their difference is small.
$\phi_{K^-}$ is skewed towards large $x$ since its valence $s$ quark is heavier than its valence $\bar{u}$ quark. However, the distribution outside the region $x\in [0,1]$ is still quite sizable (though shrinking when $P_z$ is increased).
Given that the DA for {the} largest 2 momenta
are already quite close to each other, it seems unlikely that
the residual effect in the unphysical region is totally due to higher-twist power corrections in $1/P_z$ that are not accounted for.
Given the large one-loop matching correction seen in the left-hand side of
 Fig.~\ref{fig:Kaon}, it will be important to investigate the two-loop matching contributions in the future to check their size. In addition, the truncation of $zP_z$ in the Fourier transformation can {yield}
nonzero distribution outside  $x\in [0,1]$. {This has become more visible in this work than our previous work in the pion case because
 of the larger momentum reach. }

\begin{figure}[tbp]
\centering
\includegraphics[width=.75\textwidth,]{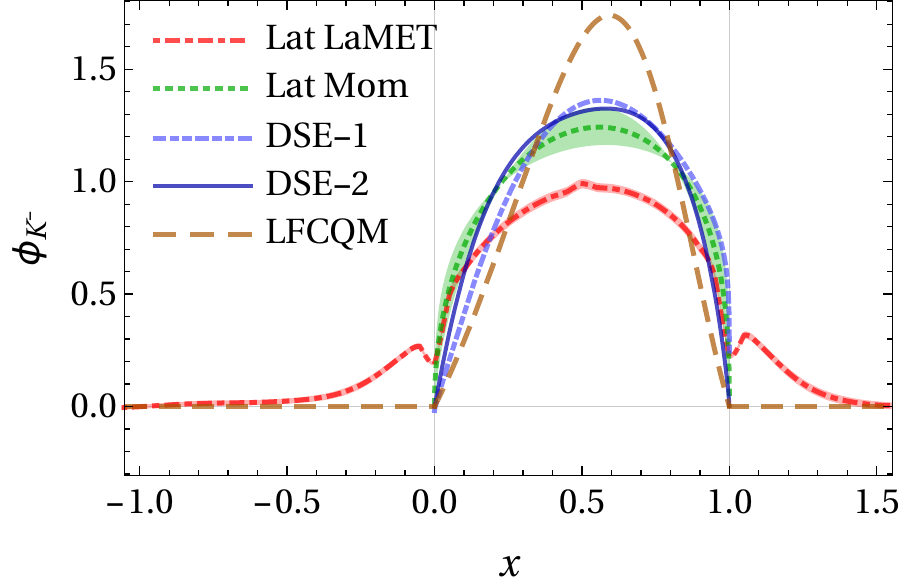}
\caption{\label{fig:Kaoncomp}
Comparison of $\phi_{K^-}$ of this work (Lat LaMET) to a few selected works in literature. This includes a parametrized fit to lattice moments (Lat Mom)~\cite{Arthur:2010xf,Segovia:2013eca},  the Dyson-
Schwinger equation calculation (DSE-1 \& -2)~\cite{Chang:2013pq}, and a light-front constituent quark model (LFCQM)~\cite{deMelo:2015yxk}. A broader distribution than the one predicted in LFCQM is clearly preferred; further studies are planned
to investigate the nonzero distribution outside $x \in [0,1]$.
}
\end{figure}

Finally, we compare our $\phi_{K^-}$ result (labeled ``Lat LaMET'') with $P_z=8 \pi/L$ with a few selected results in the literature in Fig.~\ref{fig:Kaoncomp}: the result from fitting a parametrization to the lowest few moments calculated in lattice QCD~\cite{Arthur:2010xf,Segovia:2013eca} with pion mass ranging 330--670~MeV (labeled ``Lat Mom''),
Dyson-Schwinger equation calculations~\cite{Shi:2014uwa} (``DSE-1'' and 2),
and a calculation with a light-front constituent quark model~\cite{deMelo:2015yxk} (``LFCQM'').
%
We observe a broader distribution than the one from LFCQM, without making the assumption on the distribution form of $x^\alpha(1-x)^\beta$.
Our $\phi_{K^-}$ noticeably has smaller peak near $x=0.5$; this is mainly due to the sizable distribution outside the $[0,1]$ region, {since the integral of the kaon DA is normalized to $1$.} Therefore, the DA has to have {a} smaller peak to produce the same integral. We plan to study the higher-loop matching as well as {go to} large $P_z$ to {reduce} the Fourier-transformation truncation effects.

\subsection{SU(3) Symmetry in Meson Distribution Amplitudes}
\label{subsec:su3}

\begin{figure}[tbp]
\centering 
\includegraphics[width=.48\textwidth,]{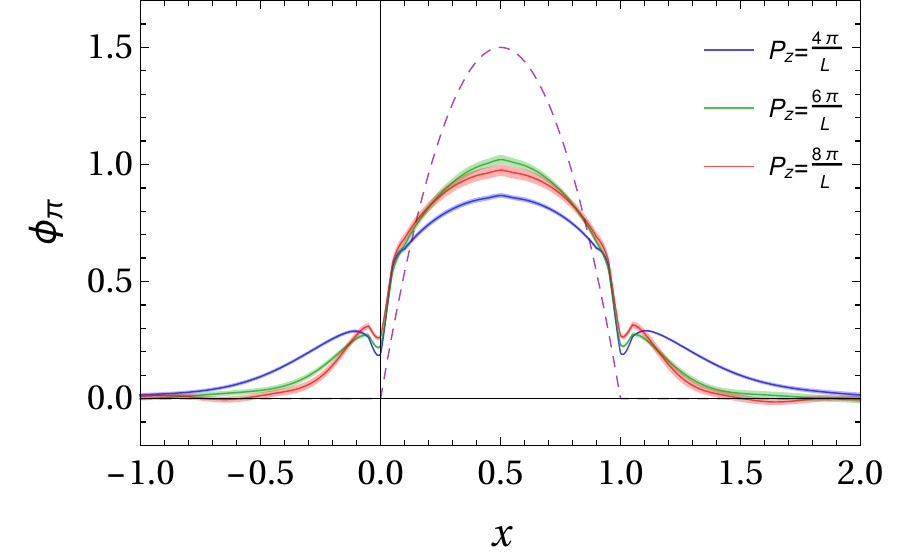}
\includegraphics[width=.48\textwidth,]{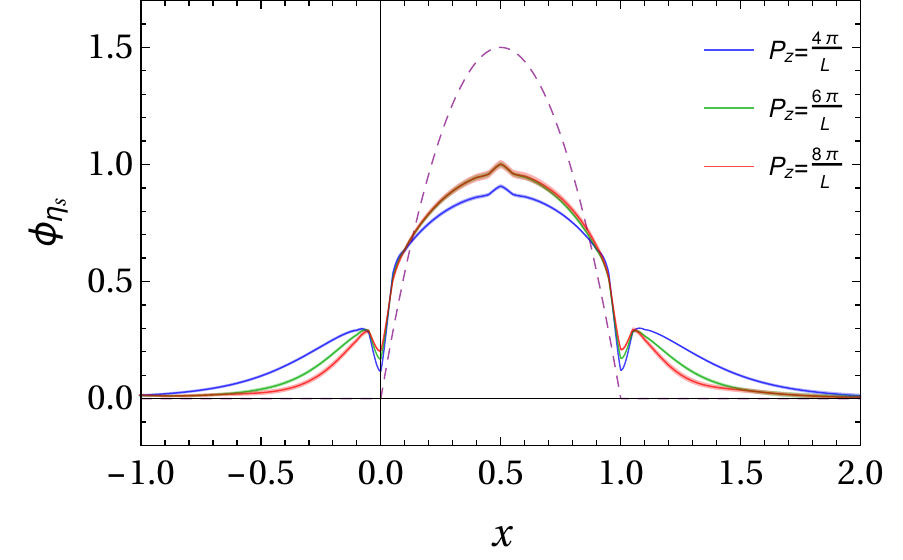}
\caption{\label{fig:PionEtasDAs} DAs of $\pi$ (left) and  $\eta_s$ (right) as functions of Bjorken-$x$ after the one-loop matching and mass corrections with different $P_z$.
The errors in the lattice calculation are statistical errors only.
The asymptotic form $\phi(x)=6x(1-x)$ is shown as the purple dashed lines. }
\end{figure}

In this work, we also update our previous study~\cite{Zhang:2017bzy} of the pion DA and make the first study of the {(connected diagrams only)} $\eta_s$ case.
Fig.~\ref{fig:PionEtasDAs} shows both DAs
obtained after the one-loop matching and mass corrections (but not higher-twist corrections at $\mathcal{O}(\Lambda_\text{QCD}^2/P_z^2)$). Larger boost momentum (with specifically tuned momentum-smearing parameters) and higher statistics are used in this work for $\phi_\pi$. The dominant systematic uncertainty, due to the counterterm $\delta m$ using a single spacing in the previous study, is significantly improved with the use of 3 lattice-spacing determinations in this work.
We also obverse that both $\phi_\pi$ and $\phi_{\eta_s}$ are symmetric with respect to $x=1/2$ due to charge-conjugation symmetry. As in the kaon case in the previous subsections, there is sizable distribution outside $x \in [0,1]$. As discussed earlier, we suspect finer lattice spacing and higher-loop matching in future studies may improve these properties.

\begin{figure}[tbp]
\centering
\includegraphics[width=.475\textwidth,]{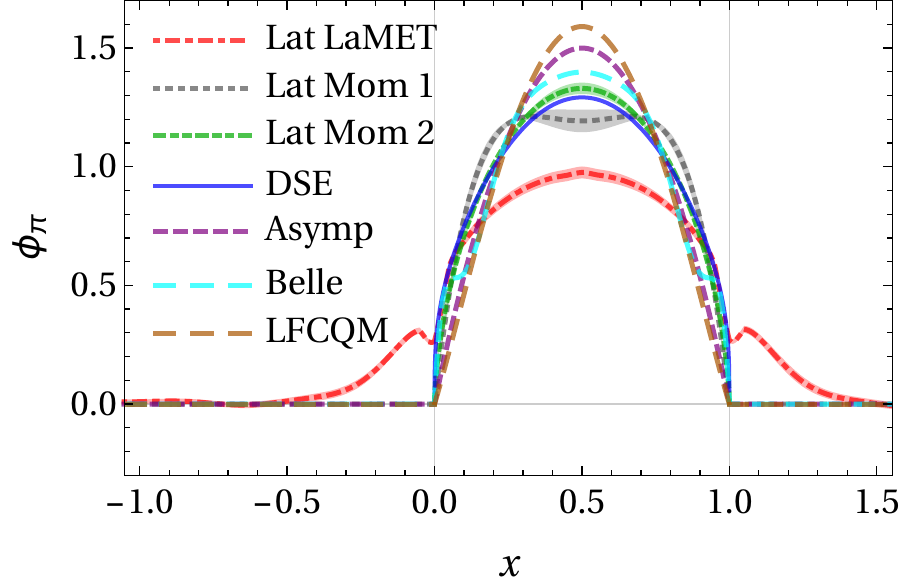}
\includegraphics[width=.5\textwidth,]{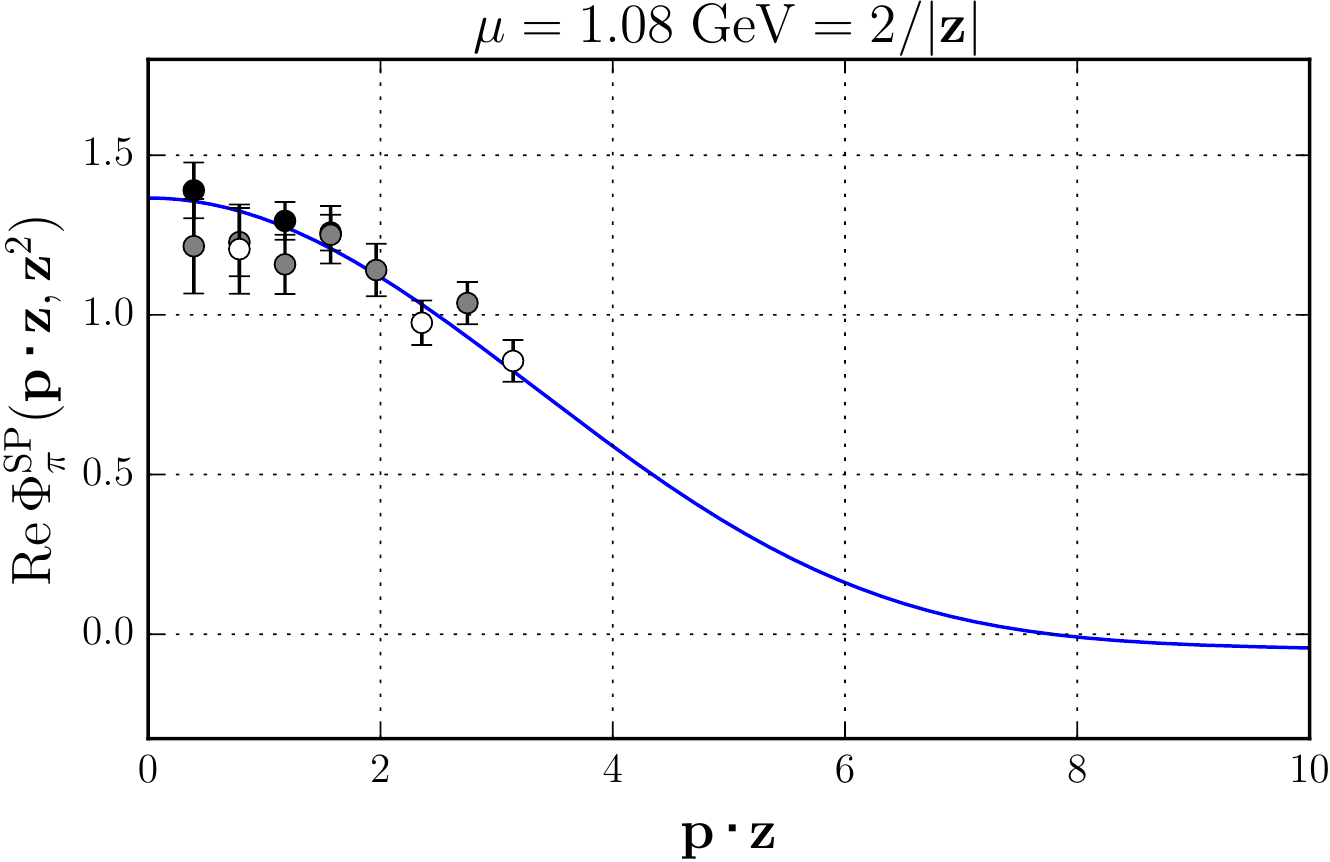}
\caption{\label{fig:PionComp}
Comparison of $\phi_\pi$ from this work (``Lat LaMET'') to previous determinations in literature. In the left panel, this includes the results from parametrized fits to the lattice moments (``Lat Mom 1'' and ``Lat Mom 2'')~\cite{Braun:2015axa}, a calculation from the DSE analysis (DSE)~\cite{Chang:2013pq}, one from the LFCQM (LFCQM)~\cite{deMelo:2015yxk}, a fit to the Belle data (Belle)~\cite{Agaev:2012tm}, and the asymptotic form $6x(1-x)$ (Asymp). In the right panel, we have converted our result on $\phi_\pi$ to the prediction for the scalar-pseudoscalar current correlator (blue curve), and compared with the lattice data for the same correlator in Ref.~\cite{Bali:2017gfr} (dark, gray and white circles, which correspond to $|\vec P|=1.08, 1.53, 1.88$ GeV, respectively, $\mu$ is the renormalization scale).
}
\end{figure}

{Figure~\ref{fig:PionComp} shows a comparison of our current results for $\phi_\pi$ with earlier results in literature. In the left panel, we show our $\phi_\pi$ result along with} a result using Dyson-Schwinger equation (DSE)~\cite{Chang:2013pq},
truncated Gegenbauer expansion fit to the Belle data for the $\gamma\gamma^*\to\pi^0$ form factor (Belle)~\cite{Agaev:2012tm},
and from parametrizations to lattice-QCD lowest-moment calculations~\cite{Braun:2015axa} to extract the pion DA.
For the fit to the Belle data, the Gegenbauer-polynomial expansion up to the eighth moment given in Ref.~\cite{Agaev:2012tm} was used at scale 2~GeV.
For the fit to the lattice-moment distribution, two different parametrizations are shown here. The first one is a simple truncation of the Gegenbauer-polynomial expansion to the pion DA parametrization to the second order $\phi(x)=6x(1-x)[1+a_2 C_2^{3/2}(2x-1)]$ (``Lat Mom 1'') with the value of $a_2$ taken from Ref.~\cite{Braun:2015axa}. The second distribution is $\phi(x)=A[x(1-x)]^B$ with $A$ and $B$ determined from the normalization condition and the lattice calculations of the second moment (``Lat Mom 2'')~\cite{Braun:2015axa}.
The two parametrizations using lattice moment calculations yield significantly different pion DA.
The difference between them can be viewed as a rough estimate of {errors} due to the moment truncation. With more lattice moment data the parametrization dependence may improve; however, with individual distributions the systematic {error} is currently underestimated.
Our distribution has a lower peak at $x=1/2$ mainly due to the nonvanishing contribution outside the $[0,1]$ region, since the integral of the distribution over all regions is normalized to 1 by definition.
Given the smallness of the mass corrections and that our curves at $P_z=6 \pi/L$ and $8 \pi/L$ are very close to each other, we expect the higher-order matching kernel will play an important role in reducing the contribution in the unphysical region. {Also higher boosted momentum will help} improve the truncation systematics in Fourier transform in $zP_z$. This needs to be further investigated before we can draw a definite conclusion on the shape of $\phi_\pi$. {In the right panel, we also compare our result on $\phi_\pi$ with the calculation using Euclidean current correlators in Ref.~\cite{Bali:2017gfr}, where the lattice data was presented for the scalar-pseudoscalar current correlator. 
In order to make a direct comparison, we have used our result to convolute with the coefficient function up to $\mathcal O(\alpha_s)$ in Ref.~\cite{Bali:2017gfr}, and then included the higher-twist contributions obtained there. Our final result is shown as the blue curve. The dark, gray and white circles are the lattice data in Ref.~\cite{Bali:2017gfr} for $|\vec P|=1.08, 1.53, 1.88$ GeV, respectively, $\mu$ is the renormalization scale. As can be seen from the plot, both approaches yield consistent results at small $\vec P\cdot \vec z$}.

Finally, we investigate the leading SU(3) flavor breaking effect predicted using ChPT.
In this work, we did not calculate $\phi_\eta$ directly; as discussed in Sec.~\ref{subsec:eta}, at the largest boost momentum $P_z=8 \pi/L$ we can approximate $\phi_\eta$ from $\phi_\pi$ and $\phi_{\eta_s}$:
\begin{align}
\label{eq:eta}
 \phi_\eta=(\phi_\pi + 2\phi_{\eta_s})/3 .
\end{align}
Since $\phi_\pi$ and $\phi_{\eta_s}$ are quite close to each other, $\phi_\eta$ is similar to the distribution shown in Fig.~\ref{fig:PionEtasDAs}.
We are interested in verifying the following SU(3) symmetry breaking relations:
\begin{align}
\label{eq:su3}
 \delta_{\text{SU}(3),1} = (\phi_{K^-}-\phi_{K^+})/2, \\
 \delta_{\text{SU}(3),2} = (\phi_{\pi}+3\phi_{\eta}- 2\phi_{K^+}-2\phi_{K^-})/8 .
\end{align}
ChPT~\cite{Chen:2003fp} predicts the magnitude of $\delta_{\text{SU}(3),1}$ to be $\mathcal{O}(m_q)$ while the magnitude of  $\delta_{\text{SU}(3),2}$ is $\mathcal{O}(m_q^2)$; thus, the lattice results should see $\delta_{\text{SU}(3),1} > \delta_{\text{SU}(3),2}$.

\begin{figure}[tbp]
\centering 
\includegraphics[width=.48\textwidth,]{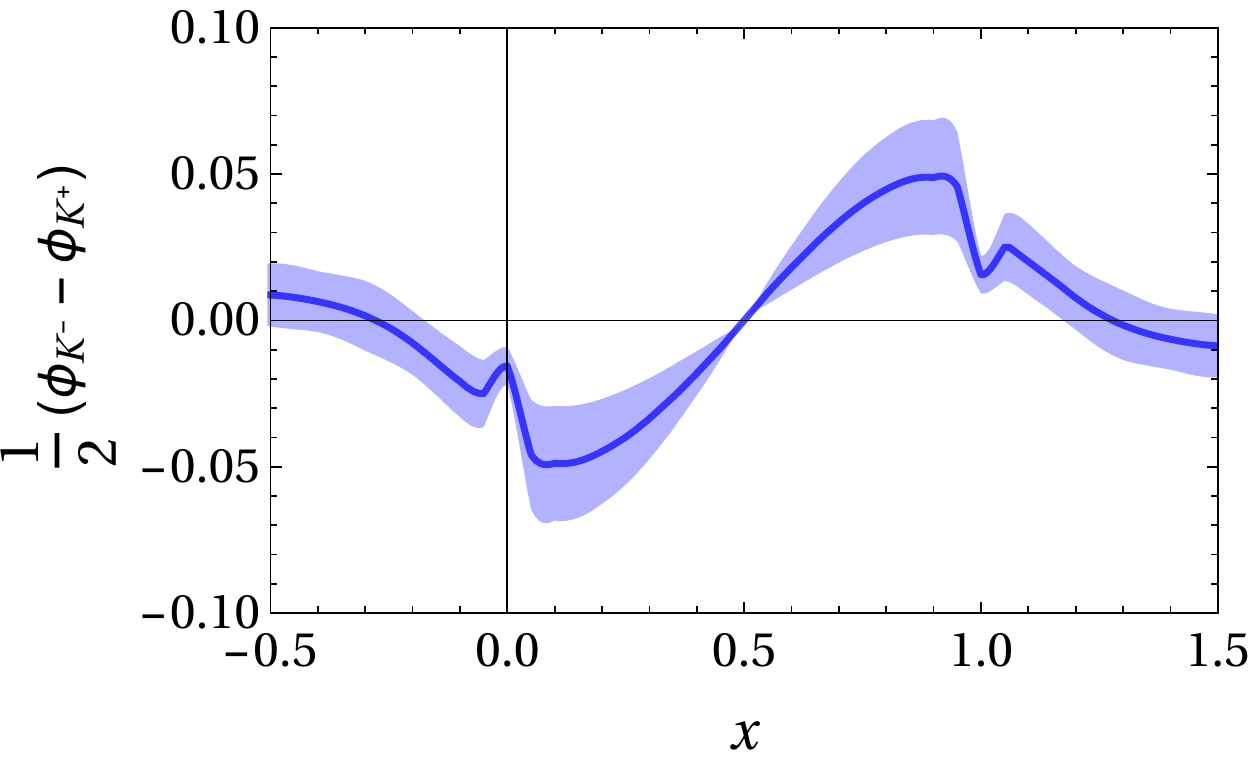}
\includegraphics[width=.48\textwidth,]{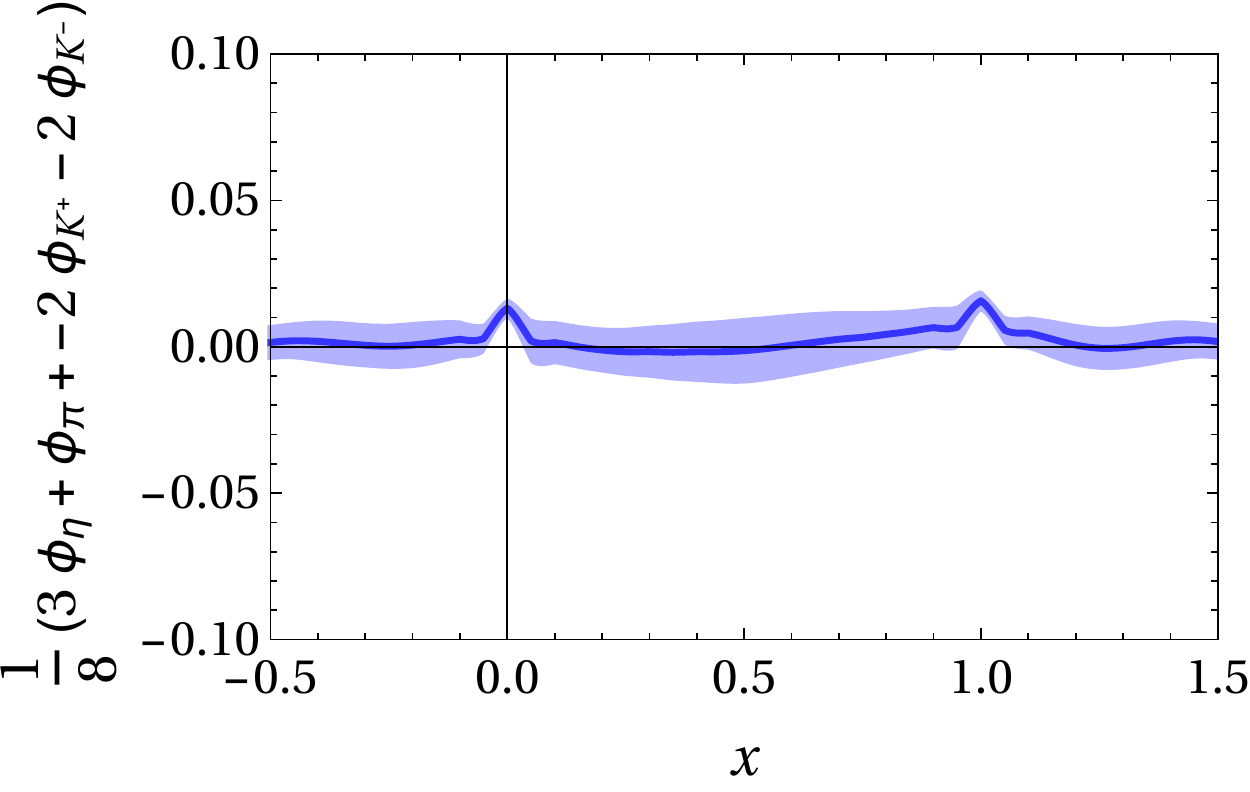}
\caption{\label{fig:SU3breaking}
Results for flavor SU(3) symmetry breaking: $\delta_{\text{SU}(3),1} =(\phi_{K^-}-\phi_{K^+})/2$ (left)
and  $\delta_{\text{SU}(3),2} = (\phi_{\pi}+3\phi_{\eta}- 2\phi_{K^+}-2\phi_{K^-})/8$ (right)
using the corrected distribution of $P_z=8 \pi/L$.
Our results support the ChPT~\cite{Chen:2003fp} prediction $\delta_{\text{SU}(3),1} > \delta_{\text{SU}(3),2}$.
}
\end{figure}

Fig.~\ref{fig:SU3breaking} shows the Bjorken-$x$ dependence of both $\delta_{\text{SU}(3),1}$ (left) and $\delta_{\text{SU}(3),2}$ (right) at the largest boost momentum $P_z=8 \pi/L$.
$\delta_{\text{SU}(3),1}$ shows a clear sign of the skewness in the kaon.
$\delta_{\text{SU}(3),1} > \delta_{\text{SU}(3),2}$ within $x \in [0,1]$ {(except when $x$ is close to $1/2$ where $\delta_{\text{SU}(3),1}=0$)}, so the {ChPT} prediction is {indeed} supported by our lattice study. In addition, $\delta_{\text{SU}(3),2}$ is consistent with zero within the statistical errors at {the pion mass of} 310~MeV. Future studies at lighter pion mass can check the quark-mass dependence directly.

\section{Conclusion and Outlook}
\label{sec:concl}
We have presented the first lattice calculation of the kaon distribution amplitude using the large-momentum effective theory (LaMET) approach with a pion mass of 310~MeV. Momentum smearing has been implemented to improve signals up to meson momentum 1.7~GeV.  We subtract the power divergence due to Wilson line using the counterterm $\delta m$ determined to $2\%$ accuracy using multiple lattice spacings{---a significant improvement over our previous pion-DA work.}
We clearly see {the} skewness of kaon from the {asymmetric} distribution {with} respect to $x=1/2$ (or equivalently the nonvanishing $\phi_{K^-}-\phi_{K^+}$). 

We also present the first results on $\eta_s$ DA (and an indirect determination of $\eta$), as well as an improved determination of the pion DAs. {Similar to the kaon case, there are non-vanishing contributions outside the physical region $[0, 1]$. Without eliminating them, we are unable to draw a definite conclusion on the shape of the DAs, since the result in the physical region will be affected by the total normalization.}
With all 3 meson DAs, we are able to investigate the leading SU(3) flavor symmetry breaking in meson DAs suggested by ChPT~\cite{Chen:2003fp}, and clearly observe $\delta_{\text{SU}(3),1} > \delta_{\text{SU}(3),2}$ for $x \in [0,1]$ {except when $x$ is close to $1/2$ where $\delta_{\text{SU}(3),1}=0$}. The quark-mass dependence can be studied in the future using lighter pion masses.

Given that some of these exciting results are being studied first time on the lattice,
there are possible improvements for future work.
With the improved signal due to the usage of the momentum-smearing sources and better determination of Wilson-loop counterterm $\delta m$,
the distribution outside the $x \in [0,1]$ region remains sizable and not consistent with zero by a few standard deviations. This leads to a few possible directions to achieve more reliable meson DAs (removing the residual DA outside the $[0,1]$ region):
Doubling the momentum on finer lattice spacing, say $0.06$~fm, can reduce the systematics due to the truncation in $zP_z$ in Fourier transform from lattice nonlocal matrix elements.
This will also reduce the size of higher-twist contributions, which seems to be more noticeable outside $x \in [0,1]$ than within.
In addition, the finite meson-momentum correction using the one-loop matching kernel dominates the sums of all corrections (including the mass correction and the higher-twist estimation).
This suggests that moving to higher-loop level for the matching kernel can have sizable contribution. We plan to work out the exact form in a future study.

\section*{Acknowledgments}
We thank the MILC Collaboration for sharing the lattices used to perform this study. The LQCD calculations were performed using the Chroma software suite~\cite{Edwards:2004sx} and multigrid solver for clover fermions package~\cite{Osborn:2010mb}. This research used resources of the National Energy Research Scientific Computing Center, a DOE Office of Science User Facility supported by the Office of Science of the U.S. Department of Energy under Contract No.~DE-AC02-05CH11231; facilities of the USQCD Collaboration, which are funded by the Office of Science of the U.S. Department of Energy,and supported in part by Michigan State University through computational resources provided by the Institute for Cyber-Enabled Research.
JWC is partly supported by the Ministry of Science and Technology, Taiwan, under Grant No. 105-2112-M-002-017-MY3 and the Kenda Foundation. LCJ is supported by the Department of Energy, Laboratory Directed Research and Development (LDRD) funding of BNL, under contract DE-EC0012704. The work of HL and YY is supposed by US National Science Foundation under grant PHY 1653405 ``CAREER: Constraining Parton Distribution Functions for New-Physics Searches''. AS and JHZ are supported by the SFB/TRR-55 grant ``Hadron Physics from Lattice QCD'', JHZ is also supported by a grant from National Science Foundation of China (No.~11405104). JHZ thanks G. Bali and P. Wein for help with the comparison plot in the right panel of Fig.~\ref{fig:PionComp}. YZ is supported by the U.S. Department of Energy, Office of Science, Office of Nuclear Physics, from DE-SC0011090 and within the framework of the TMD Topical Collaboration.

\section*{Appendix: Skewness of the Kaon Distribution Amplitude}

\begin{figure}[h]
\centering 
\includegraphics[width=.48\textwidth,]{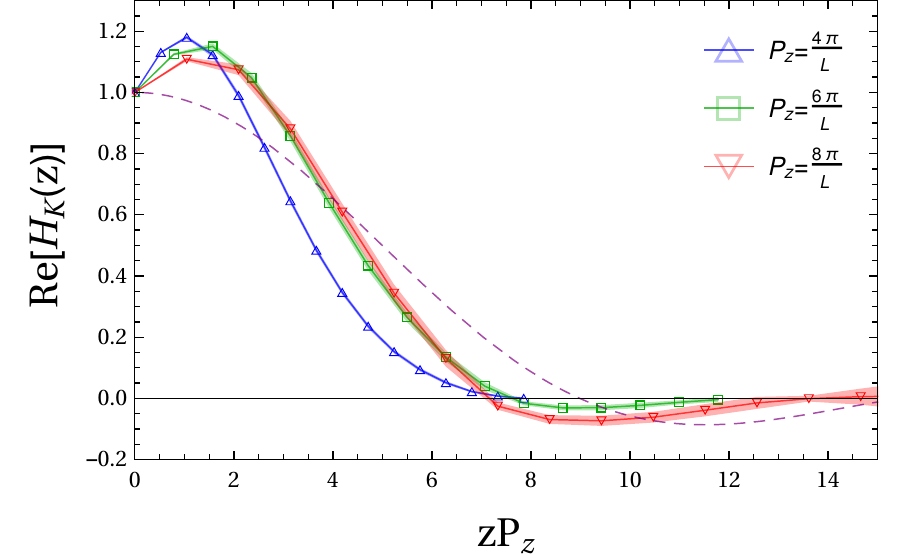}
\includegraphics[width=.48\textwidth,]{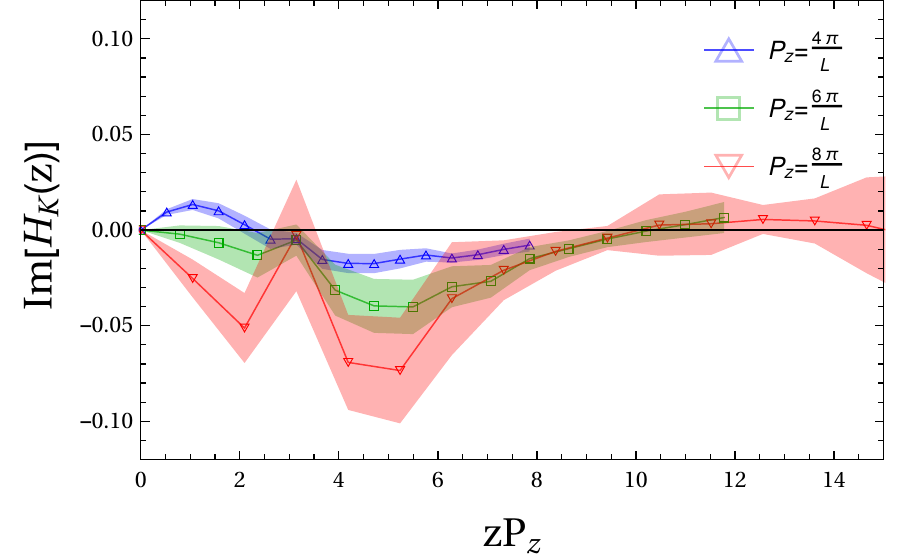}
\includegraphics[width=.48\textwidth,]{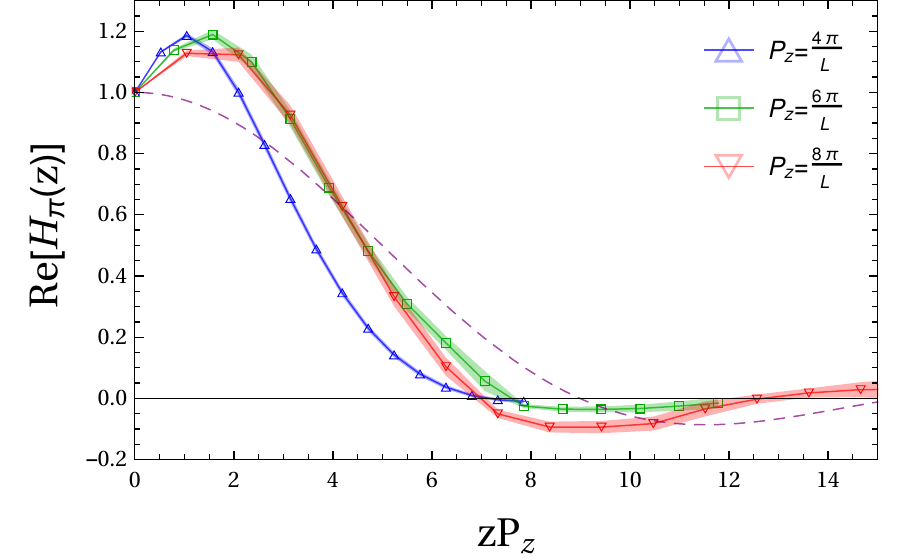}
\includegraphics[width=.48\textwidth,]{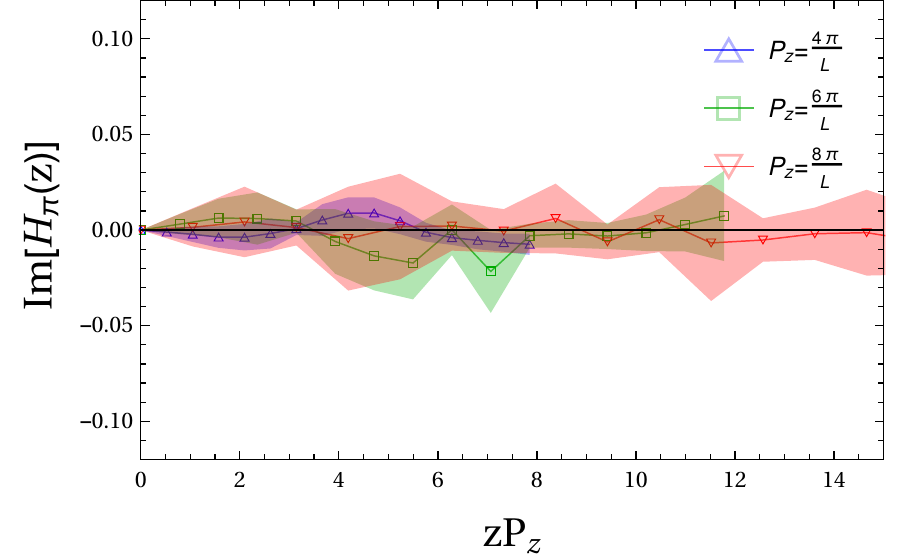}
\includegraphics[width=.48\textwidth,]{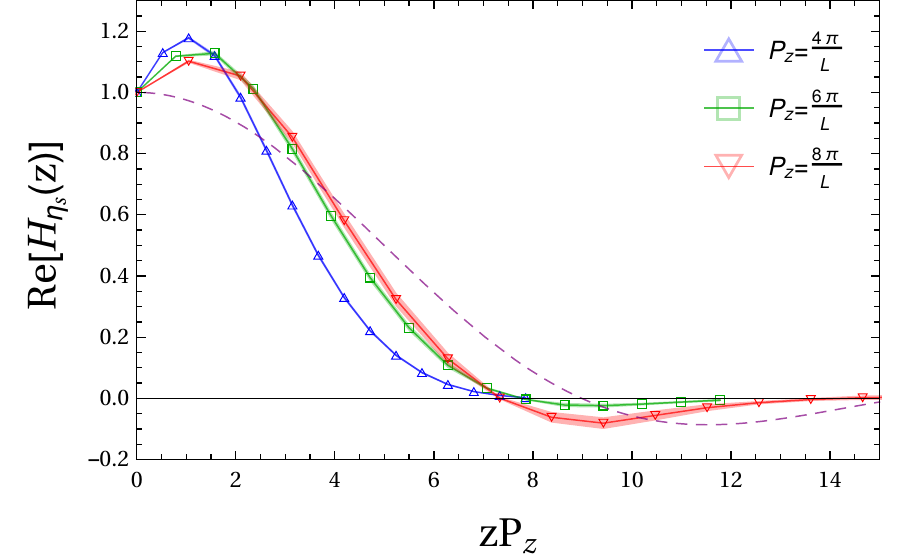}
\includegraphics[width=.48\textwidth,]{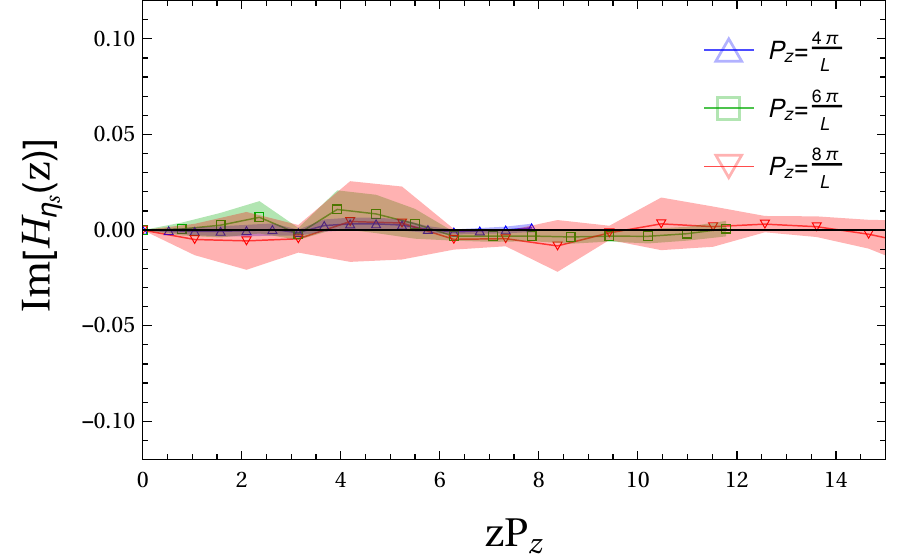}
\caption{\label{fig:improvedQuasiDAmatrix}
The improved matrix elements $H_M(z, P_z)$ defied in Eq.~\ref{eq:H} shown at three different values of $P_z$.
Since the real (imaginary) part of $H_M(z,P_z)$ is even (odd) in $z$ in definition, only the $z>0$ part is shown.
The magenta lines are derived from the asymptotic DA $\phi(x)=6x(1-x)$.
The plots in the top, middle and bottom rows are for $K^+$, $\pi$,  and $\eta_s$, respectively. The imaginary part for kaon is not vanishing, which suggests skewness in the kaon DA.
 }
\end{figure}

{There is another way to see the skewness of $\phi_K$ in the lattice data. We can consider the following improved bare matrix elements}:
\begin{align}
\label{eq:H}
H_M(z,P_z) &= e^{\frac{i}{2}z P_z + \delta m |z|} h_M(z,P_z).
\end{align}
Note that the linear divergence {has been removed by} $\delta m$.
Using $H_M(z,P_z)$ to study the skewness removes any systematics due to the truncation in $zP_z$ Fourier transform, the matching or {power} corrections.
We show {the results of $H_M(z,P_z)$} for kaon (as well as {for} pion and $\eta_s$) in Fig.~\ref{fig:improvedQuasiDAmatrix}. The real (imaginary) part of the improved meson DA matrix elements $H_K(z,P_z)$ is even (odd) in $z$; therefore, we only plot the positive-$z$ region.
The imaginary parts for $H_\pi$ and $H_{\eta_s}$ are consistent with zero due to charge-conjugation symmetry, which {make} $\tilde{\phi}^\text{imp}_{\pi,\eta_s}(x)$ and $\phi_{\pi,\eta_s}(x)$ even functions of $(x-1/2)$. On the other hand, the nonvanishing imaginary part of $H_K(z,P_z)$ leads to {an} asymmetry around $x=1/2$ and shows
 the skewness of $\phi_K$ (that is, $\phi_K(x) \ne \phi_K(1-x)$).

\providecommand{\href}[2]{#2}\begingroup\raggedright\endgroup

\end{document}